\documentclass[twocolumn]{revtex4}
\usepackage{graphicx}
\usepackage{amssymb, amsmath,amssymb,amsfonts}
\usepackage{graphicx}
\usepackage{dcolumn}
\usepackage{bm}
\usepackage{color}
\usepackage[utf8]{inputenc}
\begin{document}
\author{Francesca Di Patti}
\affiliation{Universit\`{a} degli Studi di Firenze, Dipartimento di Fisica e Astronomia and CSDC, via G. Sansone 1, 50019 Sesto Fiorentino, Firenze, Italia}
\affiliation{INFN, Sezione di Firenze, Italia}
\author{Duccio Fanelli}
\affiliation{Universit\`{a} degli Studi di Firenze, Dipartimento di Fisica e Astronomia and CSDC, via G. Sansone 1, 50019 Sesto Fiorentino, Firenze, Italia}
\affiliation{INFN, Sezione di Firenze, Italia}
\author{Francesco Piazza}
\affiliation{Universit\'{e} d'Orl\'{e}ans, Centre de Biophysique Mol\'{e}culaire, CNRS-UPR4301, Rue C. Sadron, 45071, Orl\'{e}ans, France}
\title{Optimal search strategies on complex networks}
\maketitle
{\bf \noindent Complex networks are ubiquitous in nature and play a role of paramount importance in many contexts. 
Internet and the cyberworld, which permeate our everyday life, are self-organized hierarchical graphs. 
Urban traffic flows on intricate road networks, which impact both transportation design and epidemic control. 
In the brain, neurons are cabled through heterogeneous connections, 
which support the propagation of electric signals. In all these cases,
the true challenge is to unveil the mechanisms through which specific dynamical features 
are modulated by the underlying topology of the network.  
Here, we consider agents randomly hopping along the links of a graph, with the additional 
possibility of performing long-range hops to randomly chosen disconnected nodes with a given probability. 
We show that an optimal combination of the two jump rules exists
that maximises the efficiency of target search, 
the optimum reflecting the topology of the network.}
\smallskip 

\begin{figure}[t]
\centering
\includegraphics[width=\columnwidth]{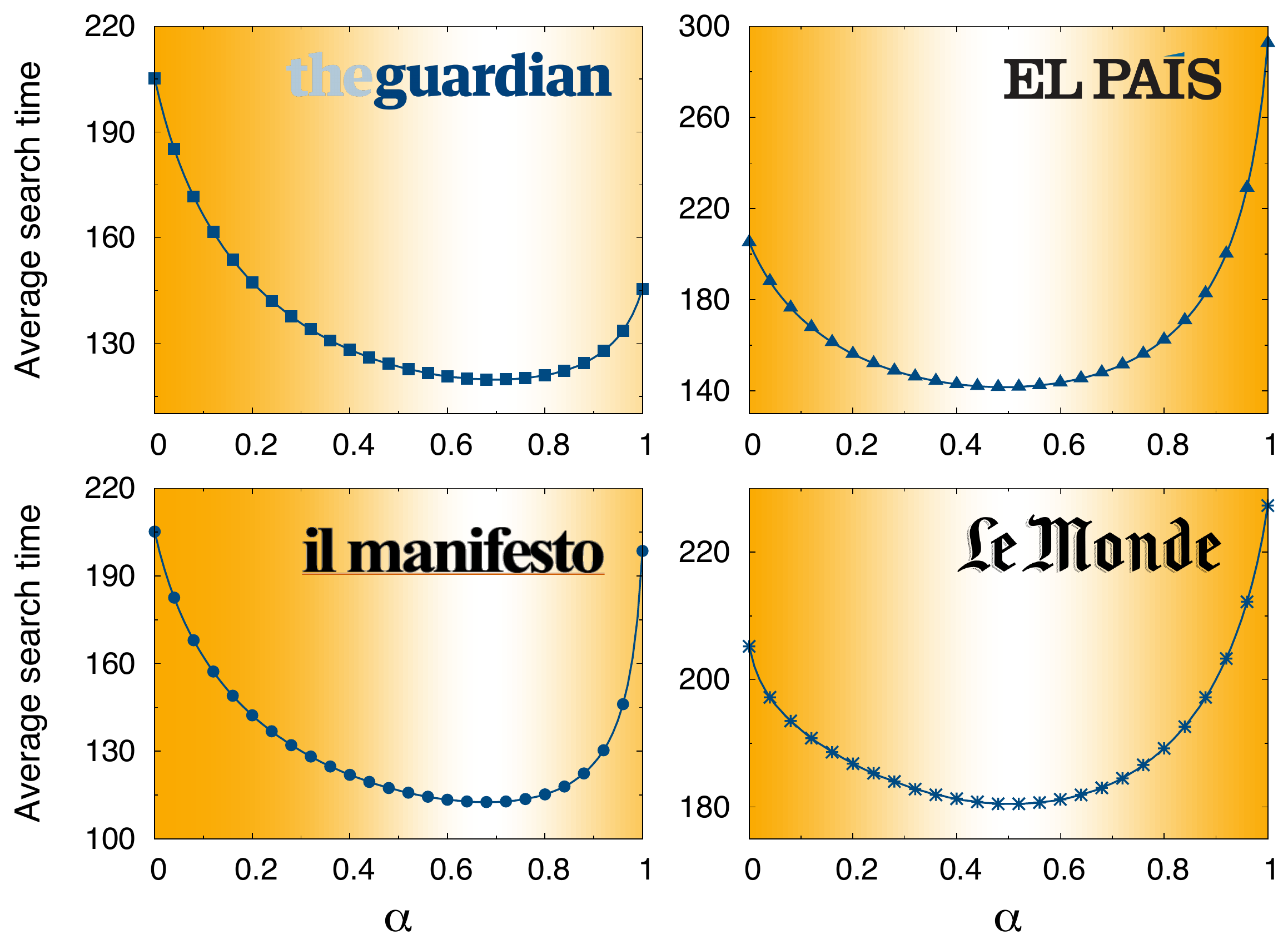}
\includegraphics[width=\columnwidth]{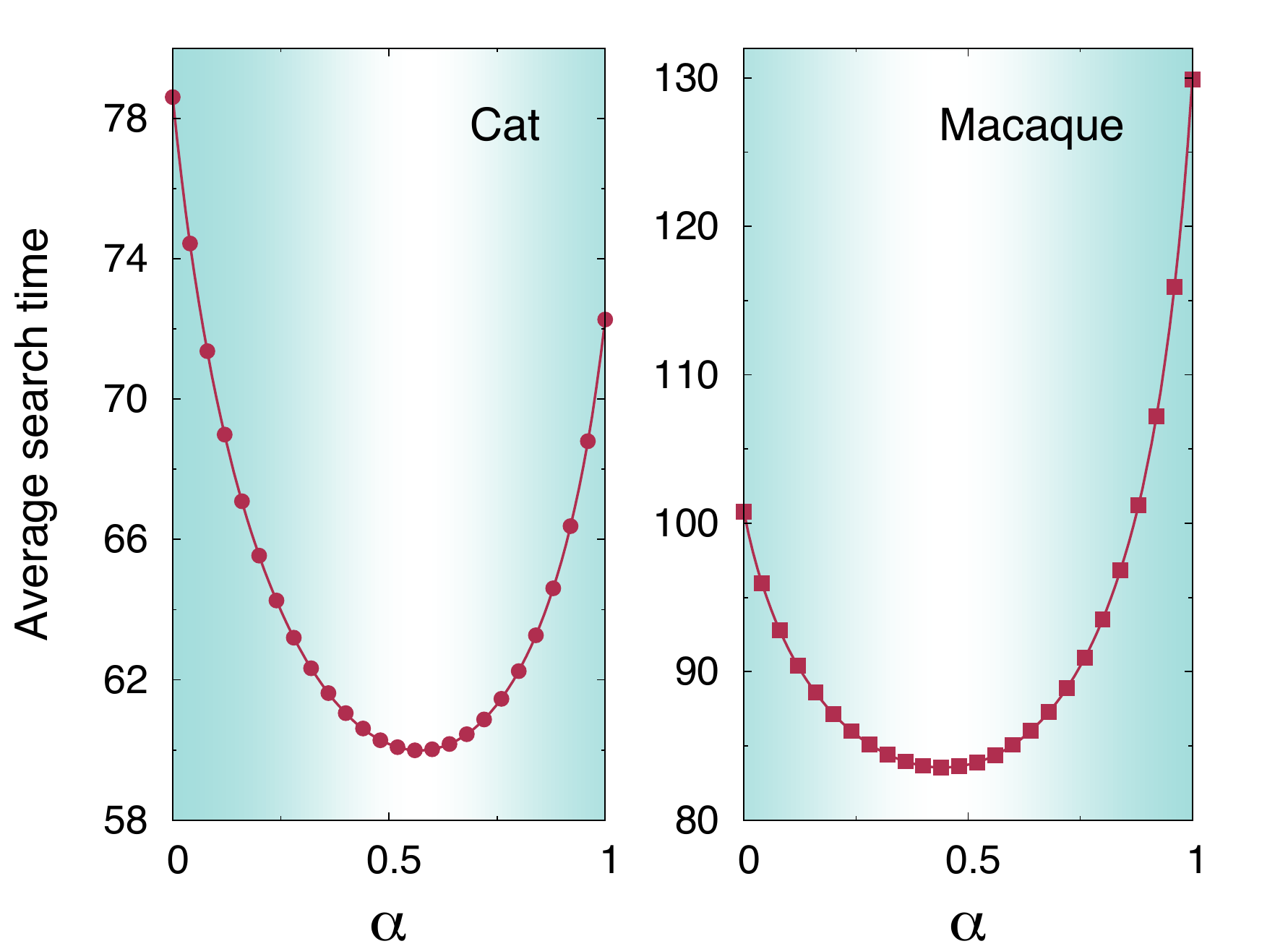}
\caption{\textbf{The average search time on real data sets displays an optimum as a function of the relative weight of 
local and long-range moves.} Top: average time $\langle t \rangle $ as a function of $\alpha$ for four real 
Web subgraphs. The $100 \times 100$ adjacency matrices have been mapped out by a Web crawler starting from the web sites 
of four major European newspapers (see Methods).  The sparse symmetric matrix $S$ has been generated with $\delta=0.04$. 
Bottom: search time in two neuronal networks: cortical connectivity network of cats  
($52$ nodes, left~\cite{Scannell:1999qo}) and macaques ($71$ nodes, right~\cite{Young:1993tw}). 
In both cases we have used $\delta=0.1$.\label{fig:natural}}
\end{figure}

\begin{figure}[t!]
\begin{center}
\includegraphics[width=\columnwidth]{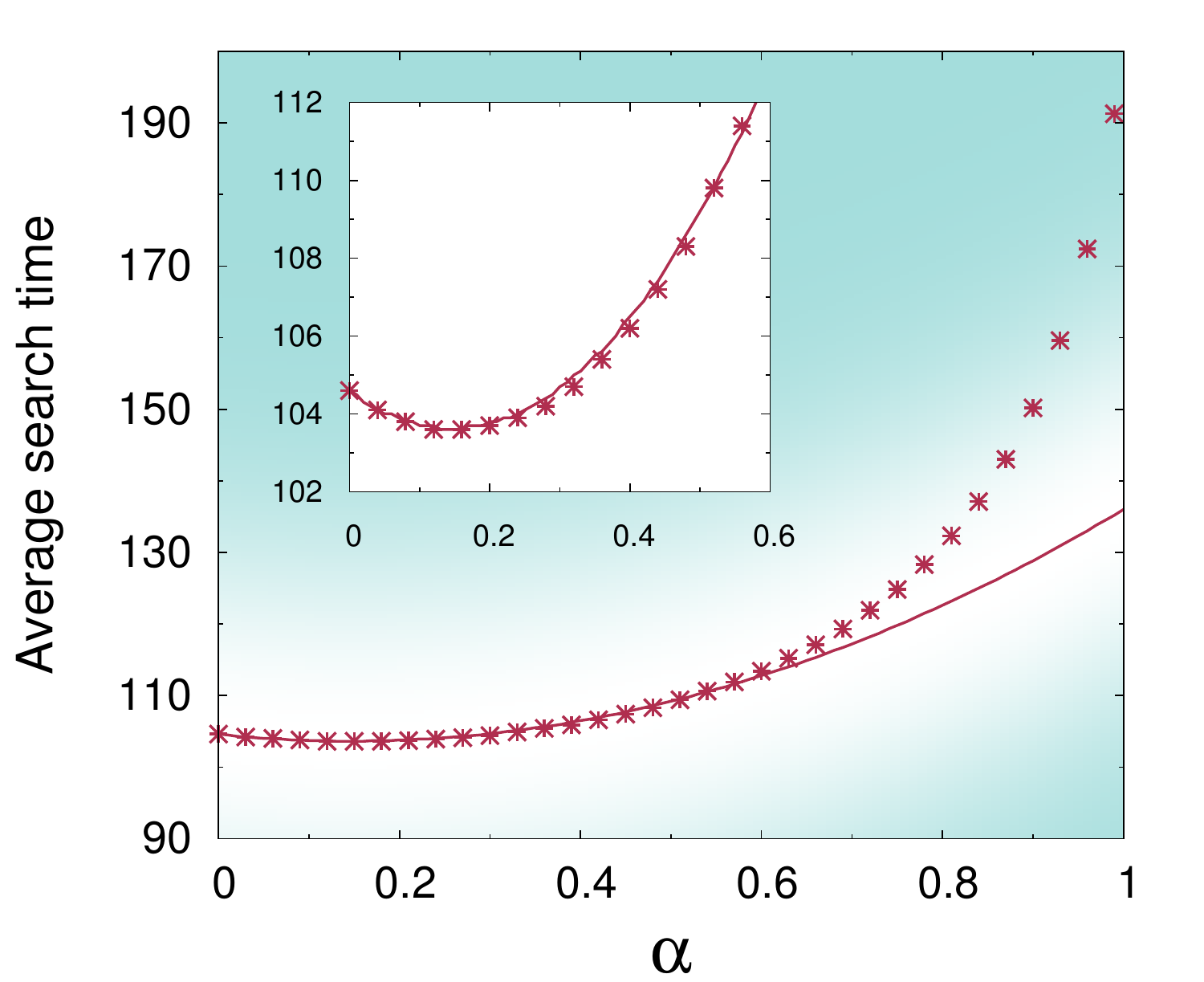}
\end{center}
\caption{\textbf{Formula~(\ref{eq:tApprox}) provides a convenient tool to enquire about the existence of 
an optimality criterion for the search time in a given network. } 
The average search time in a small random network of $N=100$ nodes 
(symbols) computed from Eq.~(\ref{eq:tMean}) is compared to the 
approximated quadratic profile described by Eq.~(\ref{eq:tApprox})
(solid line). The inset shows a close-up of the region around the minimum.  
Other parameters are: $p=0.1$, $\delta =0.29$.\label{fig:approx}}
\end{figure}

Let us consider a given agent ({\em e.g.} an electric pulse, an excitation, 
an animal or a human individual, such as a web surfer) 
located at a node of a network. The agent can hop to a neighbouring node, 
provided a link exists as specified by the adjacency matrix associated with the graph. 
The walker wanders on the network through a chain of moves, that allow for a {\it local} 
exploration of the hosting support. In such situations, the efficiency in reaching a specified 
location may be quantified by the mean first passage time, 
a robust and widely used measure of transport efficiency on networks in many contexts~\cite{Vespignani:2008hc,Baronchelli:2006ij}, 
from biology~\cite{Gallos:2007uq} and ecology~\cite{Benichou:2005kx,Benichou:2011vn,Shlesinger:2006ys} to
road network dynamics~\cite{Crisostomi:2011fk} and quantum systems~\cite{Sanchez-Burillo:2012zr,Caruso:2010ly}.\\
\indent However, local moves are not always the best option to reach a target efficiently. For example, 
facilitated diffusion in the cell nucleus, a mix of one-dimensional gliding along the DNA and three-dimensional jumps
to adjacent DNA strands, is believed to account for the efficiency of transcription factors in 
finding their binding sites~\cite{Bauer:lh,Mirny:2009fu}. 
Analogously, inspired by the behaviour of foraging animals, it has been hypothesised that the {\em local} exploration of
a connected territory might be complemented by intermittent relocation phases in order to optimize the searching 
strategy~\cite{John-OBrien:1989ve}. 
Accordingly, the animal would venture off-track through ballistic runs from time to time, thus sampling larger portions of space. 
In such examples, the relative duration of the local and relocation stages 
may control the optimization of the dual-stage strategy~\cite{Oshanin:2009qf}. \\
\indent Walkers on complex networks could in principle rely on similar integrated 
strategies, possibly tuned to the heterogeneous nature of underlying support~\cite{Ramezanpour:2007bh}. 
Let  us consider, for example, web surfing. Starting the exploration from an arbitrary web page, 
one usually proceeds by following the hyperlinks which are therein made available. 
This is a {\em local} search, which the user abandons when she opens a new tab to look for a different, 
potentially related topic, eventually landing into another virtual compartment which will be 
again probed locally for some time. On a different level, the brain displays multi-layered architectures 
of connections that assist the finely orchestrated spatio-temporal patterns  
underlying brain function~\cite{Bassett:2014dq}. One may then speculate that electric signals can be transmitted across different
layers, thus realizing {\em de facto} long-range jumps in the overall brain connectome
between single-layer connected components. \\
\indent Building upon such ideas, we investigate here the conditions for optimal target 
searches on a generic network of $N$ nodes. 
In order to quantify search efficiency on a given network, we shall compute 
mean first passage times~\cite{Lin:2014tg,Noh:2004cr,Kittas:2008kl}, which are widely used 
to gauge search strategies in many contexts~\cite{Tejedor:2009dz,Agliari:2009fu,Haynes:2008fv,Zhang:2011qa}. 
To investigate the combined effect of local and long-range moves, we 
study a simple stochastic process which accommodates for both local diffusion and long-range relocation to disconnected sites.  
Let ${\bf A}$ denote the $N \times N$  adjacency matrix of the network, with  $A_{ij}=1$ if $i$ and $j$ 
are physically connected by a  link, and $A_{ij}=0$ otherwise. The degree of node $i$ is given 
by $k_i^A=\sum_{j}^{N} A_{ij}$.   
The probability that a particle sitting at node $i$ jumps on any other node $j$ is specified by the 
following matrix
\begin{equation}\label{eq:Tij}
T_{ij}=\alpha \frac{A_{ij}}{k^A_i}+(1-\alpha)\frac{S_{ij}}{k^S_i} 
\end{equation}
where $S_{ij}=\{0,1\}$ are the entries of a random symmetric sparse $N \times N$ matrix, 
that controls the relocation via long-range hops. The density of ones in ${\bf S}$ 
is measured by the parameter $\delta\in[0,1]$, so that the average number of 
nodes that can be reached from any node $i$ via off-network 
long-range jumps is $\langle k_i^S\rangle \equiv \langle\sum_{j}^{N} S_{ij}\rangle=N \delta$. 
The  parameter $\alpha \in [0,1]$ 
tunes the relative strength of the two competing 
mechanisms, local diffusion and random relocation. When $\alpha = 1$ the walker explores the network according to a purely 
local rule, while in the opposite limit, $\alpha = 0$, hopping towards disconnected sites are the only allowed moves. 
For $\delta=1$, the matrix ${\bf S}$  is filled with ones and ${\bf T}$ becomes the known Google matrix used 
in the PageRank Algorithm~\cite{Brin:1998mi,Langville:2006pi}.\\
%
%
%%%%%%%%%%%%%%%%%%%%%%%%%%%%%%%%%%%%%%%%%%%%%%%%%%%%%%%%%%%%%%%%%%%%%%%%%%%%%%%%%%%%%%%%%%%%%%%%%%%%%%%%%%%%%%%%%%%%%%%%%%%%%
\begin{figure}[tb]
\centering
\includegraphics[width=7.9truecm]{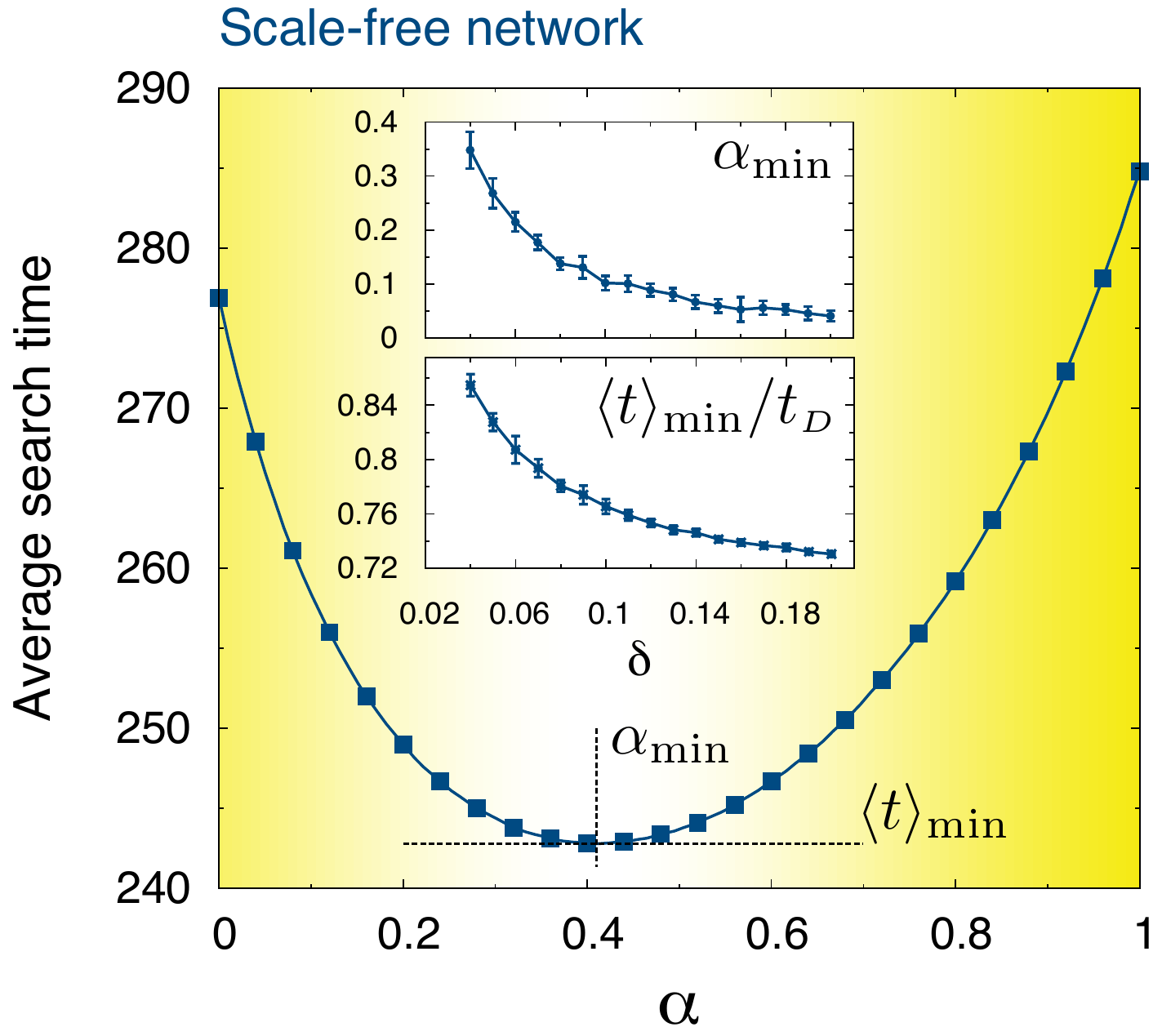}
\includegraphics[width=7.9truecm]{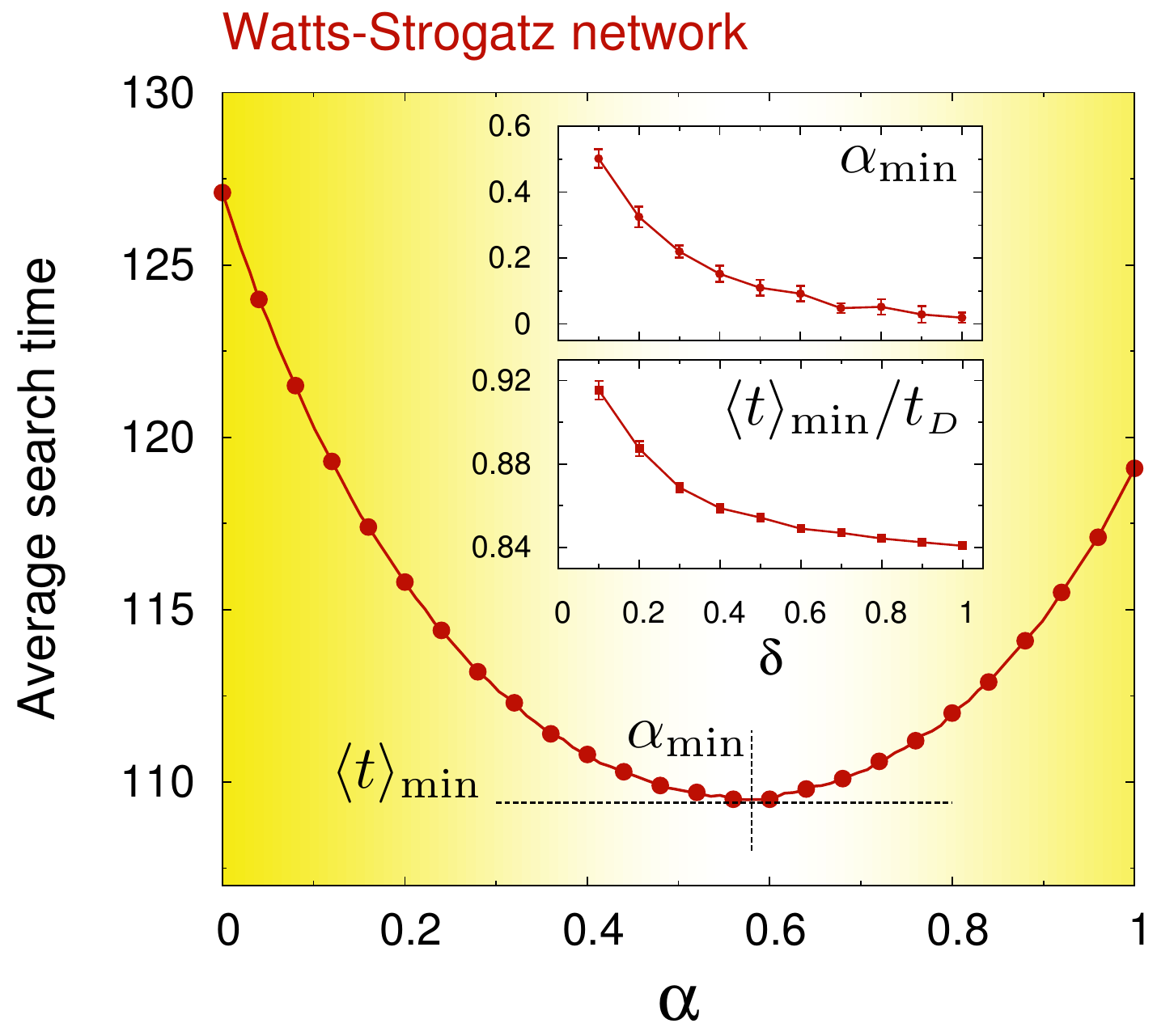}
\caption{\textbf{The average search time on synthetic networks displays an optimum as a 
function of the relative weight of local and long-range moves.} 
Upper panel: scale-free network generated with the preferential attachment method~\cite{Barabasi:1999ff} with $N=200$ and average 
connectivity $\langle k \rangle =20$. Lower panel: Watts and Strogatz small-world random 
network with $p=0.5$~\cite{Watts:1998ye}, $N=100$ and average connectivity $\langle k \rangle =9.5$. 
Here the sparse symmetric matrix $S$ has been generated 
with $\delta=0.04$ (scale-free) and $\delta=0.1$ (Watts-Strogatz). 
The insets show the position of the minimum $\alpha_{\rm min}$ and the corresponding  
shortest average time $\langle t\rangle _{\rm min}$ (normalized to the case of a purely local walker, 
$t_D \equiv \langle t\rangle_{\alpha=1}$) as a function of the average fraction of long-range 
accessible nodes $\delta$.  The data are averaged over $10$ independent realizations of the random 
matrices ${\bf S}$ and error bars correspond to one standard deviation.}
\label{fig:timeVsAlpha} 
\end{figure}
%%%%%%%%%%%%%%%%%%%%%%%%%%%%%%%%%%%%%%%%%%%%%%%%%%%%%%%%%%%%%%%%%%%%%%%%%%%%%%%%%%%%%%%%%%%%%%%%%%%%%%%%%%%%%%%%%%%%%%%%%%%%%
\indent We define the search time as the time needed by a particle starting at node $i$ to reach
an absorbing trap located at node $j$. This satisfies the following relation (see Methods)
\begin{equation}
\label{eq:tij}
t_{i \rightarrow j} =  \sum_{k=1}^{N-1}  \left ( {\bf Z}_j^{-1} \right )_{ik} 
\end{equation}
where ${\bf Z}_j = \mathbb{I}_{N-1} - {\bf T}_j$. The subscript $j$ indicates an
$(N-1) \times (N-1)$ submatrix obtained by suppressing the $j$-th row and the $j$-th column, 
while $\mathbb{I}_{N-1}$ denotes the identity matrix of size $N -1$. 
To assess the overall ability of the walker to find a target, we introduce a global 
parameter $\langle t \rangle$ by averaging Eq.~(\ref{eq:tij}) over all possible starting nodes ($i$)
and trap locations ($j$), that is,
\begin{equation}
\label{eq:tMean}
\langle t\rangle = \frac{1}{N(N-1)} \sum_{j\neq i} t_{i \rightarrow j} 
\end{equation}
In short, $\langle t\rangle$ quantifies the ability of the walker to search for targets
at the global scale of the network. The shorter  $\langle t\rangle$, the more efficient the search. The quantity 
$\alpha$ acts as a free parameter -- it can be adjusted to select the optimal balance between local  
and long-range hops, with the aim of minimizing the global exploration time.\\
\indent Fig.~\ref{fig:timeVsAlpha} illustrates how  $\langle t\rangle$ changes as a function of the relative 
weight of local and long-range moves for two different classes of synthetic undirected networks,  
the scale-free~\cite{Barabasi:1999ff,Caldarelli:2007bs} and the small-world~\cite{Watts:1998ye} networks. 
The curves display a clear minimum, implying the existence of an optimal value of $\alpha$ 
which minimizes the search time. 
Exactly the same behavior is displayed by directed networks.
The location  of the minimum depends on the topology of the 
network, which defines the backbone for local diffusion, but also on the average number of sites that can be reached 
through a single long-range hop, $N \delta$. Remarkably, the fewer sites are accessible through long-range jumps
({\em i.e.} the smaller $\delta$), the more pronounced the optimality condition (see upper 
insets in Fig.~\ref{fig:timeVsAlpha}).\\
\indent When $\delta \to 1$, $\alpha_{\rm min}$ approaches (but never reaches) 
the limiting solution $\alpha_{\rm min}=0$. In this case, the walker can virtually land on any node with just one jump 
(the matrix ${\bf S}$ is completely filled with ones), and local diffusion contributes modestly to 
further reduce the average searching time. Although a minimum always exists also for $\delta=1$ (the Google Matrix case), 
$\langle t\rangle_{\rm min}$ is very close to $N$, the time the walker needs to reach an isolated trap when $\alpha$ 
is exactly set to zero. Conversely, when $\delta <1$, long-range short-cuts are only available towards a subset of nodes. 
This is a more plausible situation, bearing in mind the afore-mentioned applications. 
When surfing the web, from time to time one will abandon a given area of exploration to look for 
the presumed central node of a new region that she wishes to sample. 
Similarly, long-range connections in the brain, established through trans-layer channels,  
are certainly fewer than those accounting for effective bridges among the $N$ nodes of a given 
layer.\\
\indent The relocation-assisted search is $10-15 \%$ more efficient  with respect to the 
purely local dynamics for intermediate values of the density $\delta$ of available distant nodes
(insets in Fig.~\ref{fig:timeVsAlpha}). The same analysis performed with different values of the average
connectivity  $\langle k\rangle$ (scale-free network) and of the rewiring parameter $p$ (Watts-Strogatz) 
yields similar results. In particular, upon decreasing $\langle k\rangle$ one 
recovers the same qualitative behaviour as obtained when increasing $\delta$ (data not shown). \\
\indent To confirm the existence of an optimal searching strategy on real data sets, 
we have extracted the adjacency graph of small portions of the web, starting from the homepages of 
four main European newspapers (see Methods). The top panel of Fig.~\ref{fig:natural} shows that the general picture 
described above for synthetic data sets is valid for real networks too. This has nothing to do with 
the peculiar structure of the Web, for the same analysis performed on small neural networks of different animals confirms 
the existence of a clear minimum in the average search time  (bottom panel in Fig.~\ref{fig:natural}). \\
%
%%%%%%%%%%%%%%%%%%%%%%%%%%%%%%%%%%%%%%%%%%%%%%%%%%%%%%%%%%%%%%%%%%%%%%%%%%%%%%%%%%%%%%%%%%%%%%%%%%%%%%%%%%%%%%%%%%%%%%%%%%%%%
%%%%%%%%%%%%%%%%%%%%%%%%%%%%%%%%%%%%%%%%%%%%%%%%%%%%%%%%%%%%%%%%%%%%%%%%%%%%%%%%%%%%%%%%%%%%%%%%%%%%%%%%%%%%%%%%%%%%%%%%%%%%%
%
\indent In all the cases examined, $\langle t\rangle $ appears to be a convex smooth function of $\alpha$ with 
a clear minimum. One may ask whether this is a widespread feature of many graphs. More generally, it would 
be helpful to have a quantitative criterion at one's disposal to predict whether an optimal 
search strategy exists  at all, possibly also identifying the optimal balance between local and long-range 
moves required to place oneself in such regime.  
Unfortunately, the exact dependence of $\langle t\rangle $ on $\alpha$ 
is hidden in the inverse of the matrix ${\bf Z}$ which, in general, cannot be  computed explicitly.
However, a criterion of this sort can be formulated by resorting to  a perturbative approach. 
If  we assume that the stationary  point is located at sufficiently small values of $\alpha$, 
then  we may consider a search time of the form
\begin{equation}
\label{eq:tApprox}
\langle t\rangle  \simeq c_0  - c_1 \alpha + c_2 \alpha^2
\end{equation}
In this case, the coefficients $c_0$, $c_1$ and $c_2$, which depend on the topology of the network, can be computed 
analytically (see Methods). A necessary and sufficient condition for a meaningful minimum to occur 
is $c_{1}>0$ and $c_{2}>0$ with $c_{1}<2c_{2}$, which ensures that $\alpha_{\rm min}<1$. 
This provides  a handy rule to enquire about the existence of an optimality condition for any given network. \\
\indent In all the cases that we examined, the coefficients $c_k$ turn out to be positive. 
Therefore  a minimum is  always predicted  to exist under the quadratic approximation, 
and closed expressions for both  $\alpha_{\rm min}$ and $\langle t\rangle_{\rm min}$ can be readily obtained. 
These match well the exact data computed through Eq.~(\ref{eq:tMean}). The agreement is of course 
better when the minimum is found close to $\alpha=0$ (Fig. \ref{fig:approx}).
%
%%%%%%%%%%%%%%%%%%%%%%%%%%%%%%%%%%%%%%%%%%%%%%%%%%%%%%%%%%%%%%%%%%%%%%%%%%%%%%%%%%%%%%%%%%%%%%%%%%%%%%%%%%%%%%%%%%%%%%%%%%%%%%%
%%%%%%%%%%%%%%%%%%%%%%%%%%%%%%%%%%%%%%%%%%%%%%%%%%%%%%%%%%%%%%%%%%%%%%%%%%%%%%%%%%%%%%%%%%%%%%%%%%%%%%%%%%%%%%%%%%%%%%%%%%%%%%%
%
Explicit analytical expressions can be obtained in some limiting cases.
When $\delta=1$ one recovers the Google matrix and the transition rate from node $i$ to node $j$ 
reads $T_{ij}=\alpha A_{ij}/k^A_i+(1-\alpha)/N$. 
In this case it is not difficult to show that $c_0= N $, $c_1= N/ (N-1)$ and 
$c_2 = N/(N-1) \left [\sum_{j=1}^N \gamma_j \left ( \gamma_j +1/k_j^A \right )  -N  \right ]$, 
where $\gamma_j=\sum_{i=1}^N A_{ij}/k_i^A$ (see Supplemental material for the full derivation). 
In the case of a regular lattice of connectivity  $c$, 
one immediately finds $\alpha_{\rm min}=c/(2 N)$ and $\langle t \rangle_{\rm min}=N-c/(4N)$. 
The more links per node are added, the larger the value of $\alpha_{\rm min}$ ($\in [1/N, 0.5]$), 
and the deeper the minimum for $\langle t \rangle$ vs $\alpha$. Although $\langle t \rangle_{\rm min}$ is shorter 
than the search time $t_{D}$ obtained for $\alpha=0$, the associated correction is just $O(1/N)$.\\
\indent Summarising, in this letter we have addressed the problem of search on networks. 
To this end, we have studied the trapping problem for a modified random walk, 
combining local hops along the links of the graph and long-range relocation jumps toward random subsets 
of disconnected nodes. 
We have shown, both for artificial and real datasets, that an optimal balance between long-range  and 
local moves exists which minimizes the average time required to reach a trap. 
Furthermore, closed analytical expressions have been derived, enabling one  to predict the optimal combination 
as a function of the network topology. The optimality criterion seems to be a universal dynamical 
mechanism, which might have exerted a critical pressure in the evolutionary selection of 
many naturally occurring network architectures and that might equally well 
be exploited in the optimization of human-made technological solutions.

\vspace{0.5 truecm}

\section*{Methods}

\noindent\paragraph*{\bf Computation of first passage times.}
The mean first passage time $t_i$, namely the time it takes for a walker 
starting at site $i$ to get to any one of $N_{\Gamma}$ randomly placed traps, can be computed by 
extending to the case of a network the standard argument used in the continuum limit for a random walk on a line. 
Let us consider the interval $[0, x_0]$ on the real axis and  a random walk with two absorbing boundaries 
located at $x = 0$ and $x = x_0$. The time interval between two jumps is $\Delta t$ and the lattice 
spacing is $\Delta x$. The exit time $t(x)$ obeys 
to $t(x)= 1/2 \left [ t(x+\Delta x) +\Delta t \right ] + 1/2 \left [ t(x-\Delta x) +\Delta t \right ]$ 
meaning that the walker can be regarded as starting one step in the future with equal probability 
from either $x+\Delta x$ or $x-\Delta x$. The generalization of this equation for a random walk on 
a network is simply given by $t_i=\sum_j T_{ij}\left [ t_j +\Delta t \right ]$, a formulation which 
proves particularly convenient to investigate the trapping problem. 
Indeed, re-labelling the nodes of the network so as to have non-trap nodes going 
from $1$ to $N -N_{\Gamma}$ and all traps located at nodes $N-N_{\Gamma}+1$ to $N$, 
one obtains a matrix ${\bf T}$ with the last $N_{\Gamma}$ rows equal to zero. 
Rearranging correspondingly the array $t_i$ and recalling that $t_i=0$ for $N-N_{\Gamma}+1 \leqslant i \leqslant N$, 
one finds that the exit times are solution of the linear system  
$\sum_{j=1}^{N-N_{\Gamma}} Z_{ij}t_j=1$, where we have denoted by ${\bf Z}$ the 
upper-left $(N-N_{\Gamma}) \times (N-N_{\Gamma})$  block square sub-matrix of 
${\bf T}-\mathbb{I}_N$. Eq. (\ref{eq:tij}) is the formal solution of this last equation.
\noindent\paragraph*{\bf Adjacency matrices of sub-networks from the web.} 
To gather real data from the Web we have used the Web crawler {\em surfer.m} (\texttt{http://www.mathworks.com/}). 
Starting from a selected URL, the crawler identifies all the hyperlinks in the page and adds them to 
the list of URLs to visit. Once all these URLs are visited, the procedure is repeated recursively for each URL in the list  
until the assigned number of websites is reached.  The outcome of the algorithm  is stored in an 
adjacency matrix where nodes represent the visited pages: the entries of the matrix are $1$ if two pages are 
connected trough a hyperlink, $0$ otherwise. The matrix is then symmetrized.
\noindent\paragraph*{\bf Perturbative expansion of a sum of two matrices.}
Let ${\bf C}$ and ${\bf B}$ be two arbitrary non-singular square matrices of the same dimension and 
let us introduce the operator $\Theta$, that returns the sum of all the elements of a given square matrix .
Starting from the relation $\left( {\bf C}+\epsilon {\bf B}\right)^{-1}=  \left( \mathbb{I}_N+\epsilon {\bf C}^{-1}{\bf B}\right) ^{-1}  {\bf C}^{-1}$, and expressing $\epsilon  {\bf C}^{-1}{\bf B}$  by a Neumann series~\cite{Stewart:1998il}, it follows $ \left({\bf C}+\epsilon {\bf B} \right )^{-1} = {\bf C}^{-1} -\epsilon {\bf C}^{-1} {\bf B} {\bf C}^{-1} + \epsilon^2 {\bf C}^{-1} {\bf B} {\bf C}^{-1} {\bf B} {\bf C}^{-1} + \ldots$.
To apply this approximation to Eq.~(\ref{eq:tMean}), we introduce two diagonal matrices 
associated with ${\bf A}$ and ${\bf S}$, namely ${\bf K_A}=\text{diag}(k_1^A, \ldots , k_{N}^A)$  
and ${\bf K_S}=\text{diag}(k_1^S, \ldots , k_{N}^S)$.  In this way, ${\bf T}$ takes the form $\alpha {\bf K_A}^{-1} {\bf A} +(1-\alpha) {\bf K_S}^{-1} S$. Consequently, by denoting again by $j$ the position of the trap, the terms of the reduced 
matrix ${\bf Z}_j$ can be easily rearranged by collecting together those proportional 
to $\alpha$. In formulae:  ${\bf Z}_j= \mathbb{I}_{N-1}-({\bf K_S}^{-1})_j {\bf S}_j +\alpha \left [  ({\bf K_S}^{-1})_j {\bf S}_j -({\bf K_A}^{-1})_j {\bf A}_j \right ]$ . Setting ${\bf C}_j= \mathbb{I}_{N-1}-({\bf K_S}^{-1})_j{\bf S}_j$, ${\bf B}_j=({\bf K_S}^{-1})_j {\bf S}_j-({\bf K_A}^{-1})_j {\bf A}_j$ and $\epsilon =\alpha$, and applying the operator $\Theta(\cdot)$ to $\left( {\bf C}_j+\epsilon {\bf B}_j\right)^{-1}$, we recover Eq. (\ref{eq:tApprox}) with $c_0= \sum_j \Theta \left ({\bf C}_j^{-1} \right ) /N/(N -1)$, $c_1= \sum_j \Theta\left ({\bf C}_j^{-1} {\bf B}_j {\bf C}_j^{-1}\right ) /N/(N -1)$ and $c_2= \sum_j\Theta\left ({\bf C}_j^{-1} {\bf B}_j {\bf C}_j^{-1} {\bf B}_j {\bf C}_j^{-1}\right ) /N/(N -1)$.

\acknowledgements The authors would like to thank Alessio Cardillo for his critical reading of the manuscript. This work has 
been partially supported by Ente Cassa di Risparmio di Firenze and program PRIN 2009 founded by the Italian Ministero dell'Istruzione, dell'Universit\`{a}  e della Ricerca (MIUR). 
%
%
%===================================================================================================================================================
% REFERENCES
%
%\bibliography{MFPT_netw_search} 
%

%
%
%
\end{document}